# Passive Phased Array Acoustic Emission Localisation via Recursive Signal-Averaged Lamb Waves with an Applied Warped Frequency Transformation

L. Pollock[1], G. Wild[1]

[1] School of Engineering and Information Technology, UNSW at the Australian Defence Force Academy, Northcott Drive, Campbell, Australian Capital Territory, 2612, Australia

**Abstract**

This work presents a concept for the localisation of Lamb waves using a Passive Phased Array (PPA). A Warped Frequency Transformation (WFT) is applied to the acquired signals using numerically determined phase velocity information to compensate for signal dispersion. Whilst powerful, uncertainty between material properties cannot completely remove dispersion and hence the close intra-element spacing of the array is leveraged to allow for the assumption that each acquired signal is a scaled, translated, and noised copy of its adjacent counterparts. Following this, a recursive signal-averaging method using artificial time-locking to denoise the acquired signals by assuming the presence of non-correlated, zero mean noise is applied. Unlike the application of bandpass filters, the signal-averaging method does not remove potentially useful frequency components. The proposed methodology is compared against a bandpass filtered approach through a parametric study. A further discussion is made regarding applications and future developments of this technique.

## Introduction

Guided Waves (GWs) have found considerable application for the interrogation of structures and materials [1, 2]. Of the numerous GWs that have been studied, Lamb waves have shown promise due to their ability to traverse over great distances and strong tendencies to interact with damage [1]. Passively acquired GWs - or Acoustic Emissions (AEs) - occur natively in a structure as a result of a transient event, such as crack growth or impact. The release of energy from said event can be detected through suitable instrumentation and analysed to infer information regarding the event or waveguide, i.e., the structure. This paper explores the application of Passive Phased Arrays (PPAs), being a phased array that acquires an AE for the interrogation of a structure. A PPA leverages the small intra-element spacing of adjacent sensors to limit the effects of dispersion in Lamb waves that can lead to difficulties in analysis.

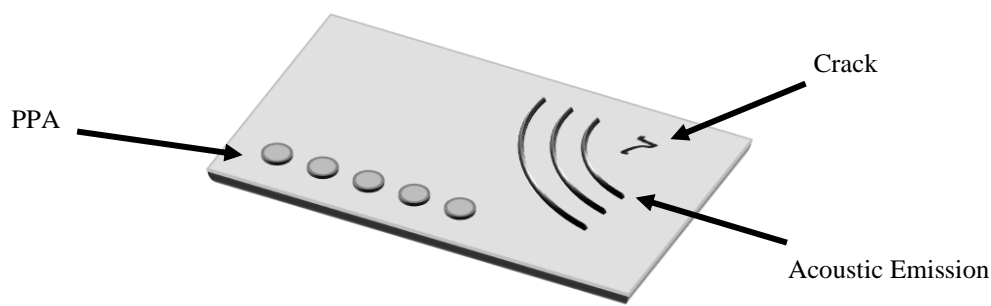

*Fig. 1: A PPA setup to detect an AE from crack growth.*

Whilst effective, a Warped Frequency Transformation (WFT) is also applied to the signals to compensate for any dispersion that may be present. The combination of the two methodologies, physical and computational, are a robust means of overcoming signal dispersion. Following this, the acquired signals are assumed to be dispersion-free, and a recursive signal-averaging method is used to remove noise before the source is localised.

## Theory

**Warped Frequency Transformation (WFT)**

The WFT is a unitary Time-Frequency Transformation (TFR) that reshapes the frequency axis through a warping function $w(f)$ to compensate for signal dispersion. Due to the time-frequency uncertainty principle, the capability of a TFR to distinguish between closely spaced modes over a wide frequency range is limited [3]. As such, the WFT was designed to allow for enhanced mode extraction due to a more flexible tiling of the time-frequency domain [3]. The warping function is designed to ensure invertibility such that the frequency axis is mapped onto itself, i.e.,

$$\dot{w}(f) > 0 \tag{1.1}$$
$$\exists\, w^{-1}, \quad w^{-1}(w(f)) = f \tag{1.2}$$

For the case of a discretely sampled signal, the modified Fourier transform is given as follows over a range $M$ of discrete frequencies: $f_k = k/M, \ k \in \{-M/2, \ldots M/2\}$ [4].

$$\mathcal{F}_{wD}[x(f_k)] = \sqrt{\dot{w}(f_k)} \sum_{n \in \mathbb{Z}} x(n) \cdot e^{-2\pi i n w(f_k)} \tag{2}$$

The warping function is determined considering an arbitrary constant $K$ that is selected such that $w^{-1}(f_0) = w(f_0) = f_0$ to ensure that the warped signal has the same peak frequency as the original signal. Hence, $K = 1/c_P(f_0)$.

$$w^{-1}(f) = \frac{f \cdot c_P(f_0)}{c_P(f)} \tag{3}$$

**Recursive Signal Averaging Method**

An advantage in the use of a PPA is the capability to assume that each acquired signal is a scaled, translated, and noised copy of its adjacent counterpart. These assumptions for two time-domain signals separated by a time delay ($\tau$) are mathematically represented as follows wherein $x$ is the measured signal, $s$ is the signal at the source, $z$ is the noise component, and $A$ is the amplitude scale factor.

$$x_1(t) = s_1(t) + z_1(t) \tag{4.1}$$
$$x_2(t) = As_1(t + \tau_{12}) + z_2(t) \tag{4.2}$$

For the case where the noise is directly sampled and is normally distributed with zero-mean and variance of $\sigma^2$, an artificial time-locking procedure may be applied in post-processing to reduce noise level. This technique is a low-level data fusion methodology that combines several raw signals to produce new raw signals. The process of signal averaging allows for a maximum reduction in the noise level by a factor of $\sqrt{n}$, where $n$ is the number of measured signals.

$$Var\left(\frac{1}{n}\sum_{i=1}^{n} z_i(t)\right) = \frac{1}{n^2} Var\left(\sum_{i=1}^{n} z_i(t)\right) = \frac{1}{n^2} \sum_{i=1}^{n} Var(z_i(t)) \tag{5}$$

Assuming that the noise variance is constant, the following may then be deduced.

$$\frac{1}{n^2}\sum_{i=1}^{n} Var(z_i(t)) = \frac{1}{n^2}n\sigma^2 = \frac{1}{n}\sigma^2 \qquad (6)$$

The aforementioned methodology is predicated on the assumption that the signals are perfectly time-locked, which is difficult to satisfy as time-locking in post-processing is ultimately dependent upon the noise itself and cannot be defined relative to some absolute trigger. Hence, a recursive routine is implemented that performs multiple time-locking procedures using the generalised cross-correlation. To accomplish this, each signal is aligned to an arbitrary time $t_0$ that functions as the pseudo trigger point. Due to the influence of dispersion upon the accuracy of the cross-correlation function, the central signal in the array is used as the reference about which the other signals are aligned. As a result, the relative dispersion between signals is minimised. Through multiple time-locking events, the aligned signals are averaged to obtain a mean representation of the noiseless signal, named the carrier signal. It is hypothesised that the performance of the recursive signal-averaging approach is highly dependent upon the number of and sampling frequency of the signals. Upsampling the signals by method of cubic spline interpolation may improve the performance of this approach but is highly computationally expensive. The recursive signal-averaging approach does not require prior knowledge of the signal structure and hence, may be more suitable for the analysis of signals in which little information is known regarding the source or waveguide. Extension of the recursive signal-averaging approach is likewise explored wherein WFT may be effectively employed to compensate for dispersion. Hence, large, distributed sensor networks observing AEs would benefit the greatest from the application of WFT followed by recursive signal-averaging.

## Methodology

To determine the usefulness of the WFT for the application of a PPA, two independent parametric studies have been undertaken. In the first study, the WFT is applied to noiseless signals and localised to compare its performance with its raw signal counterpart. The metrics of absolute Root Mean Square (RMS) position error and Direction of Arrival (DoA) have been used to assess the accuracy of the localisation. The WFT has been studied for all combinations of wave modes as well as sensor spacings up to 100 mm. The source of the AE has been fixed at a relative location of (100, 100) mm to the centre of the sensor array and a cross-correlation has been performed to estimate the time delay between adjacent signals. As the WFT is best suited to remove the dispersion of the signal, an input has been defined to maximise the dispersion within the symmetric mode. To accomplish this, the thickness of the panel is chosen as the point of inflection for the symmetric dispersion curve corresponding to a 200 kHz input. Aluminium has been chosen as the material of study with mechanical properties given in Table 1. The frequency-thickness product for this input is identified as 2,115 kHz mm, corresponding to a panel thickness of 10.575 mm. A Gaussian apodised sinusoid with a 30% fractional bandwidth has been used as the input signal for all tests. The dispersion curves for the aluminium plate are given in Fig. 2 a), illustrating the inflection point - corresponding to the peak frequency - and the upper and lower frequency bounds. The frequency representation of the input signal is likewise given in Fig. 2 b). The signals have been simulated discretely from first principles using a sampling frequency of five Million Samples Per Second (MSPS).

*Table 1: Mechanical properties of aluminium used in the study.*

| Property | Symbol | Value |
|---|---|---|
| **Young's Modulus** | $E$ | 70 GPa |
| **Density** | $\rho$ | 2780 kg/m$^3$ |
| **Poisson's Ratio** | $\nu$ | 0.33 |

The second parametric study to validate the signal-averaging methodology employs a two-step procedure. Firstly, a recursive cross-correlation upon the time-domain signals is applied to align them, following which they are averaged. A threshold number of alignments has been arbitrarily selected for validation purposes to overcome any bias in the alignment procedure. As a result of the dispersion in the symmetric mode, it is likely that alignment of the raw signals will be difficult to perform, and hence, use of the WFT should improve performance. An input signal at a location of (1000, 1000) mm has been used to ensure adequate dispersion may occur from which up to 11 sensors are used to perform the signal averaging. The sensor array is co-linear with an intra-element spacing of 6.5 mm, corresponding to approximately half the wavelength of the antisymmetric mode at 200 kHz. The Signal-to-Noise (SNR) ratio of the signals has been measured and averaged across ten different noise seeds. An SNR of ten has been applied to mimic experimental observations [5]. This initial study is used to validate the performance of the technique through use of the SNR metric. Following this, the symmetric mode was extracted from the signal-averaged WFT signal and localised using beamforming principles. Furthermore, to act as a control, the raw signal was bandpass filtered between 50 kHz and 300 kHz following which it was localised through the same procedure.

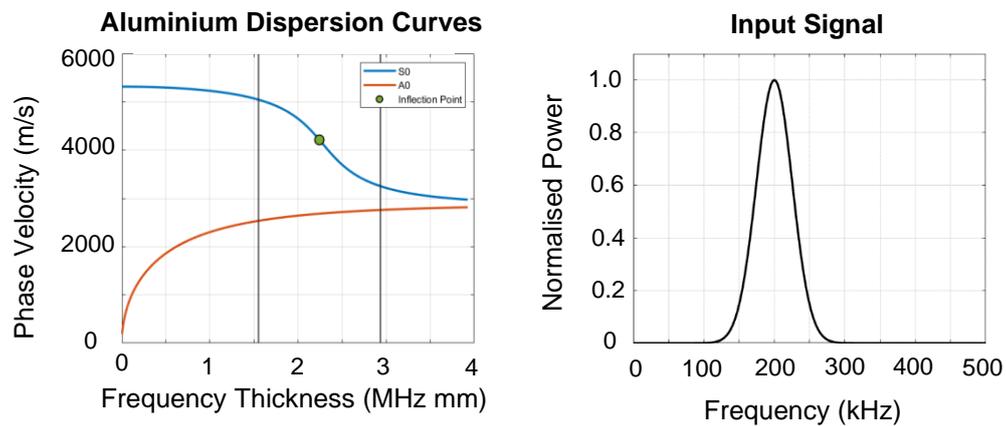

*Fig. 2: a) The dispersion curves and inflection point for the input and b) the frequency representation of the input signal.*

## Results

### WFT Validation

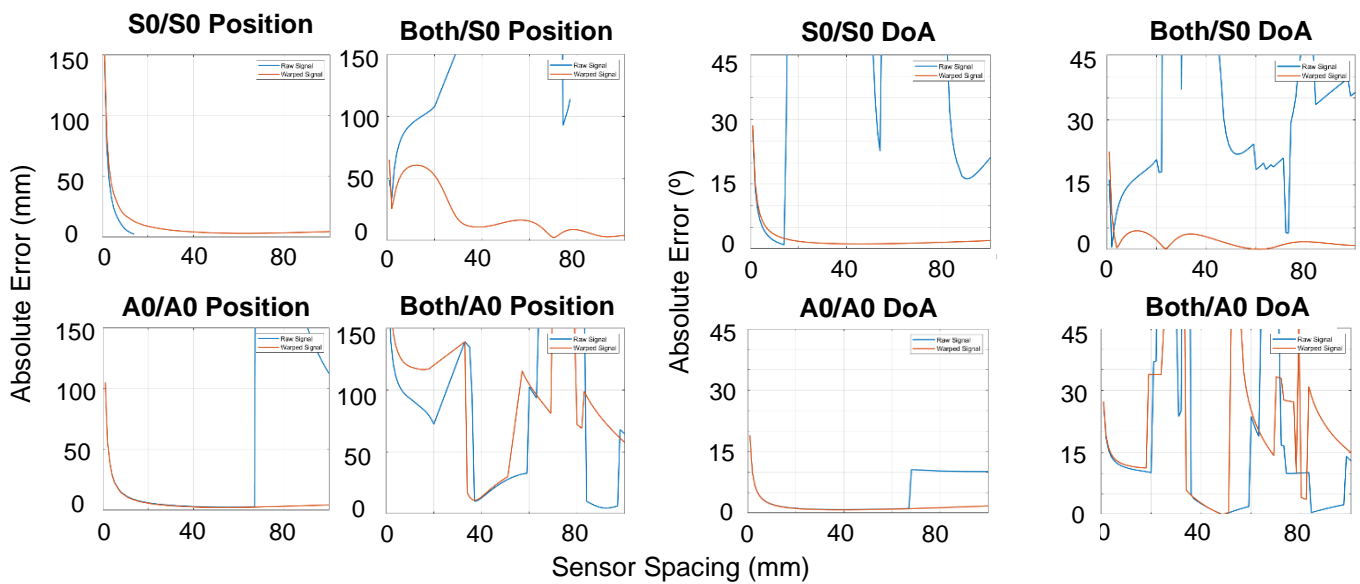

*Fig. 3: Mono-modal and bi-modal analysis results for the 200 kHz input.*

The results of the first parametric study, conducted as a verification of the WFT for dispersion compensation are highlighted in Fig. 3. The titling convention 'Lamb Mode/Warp Mode' has been applied to distinguish between the results. For example, 'S0/S0' indicates that only the symmetric mode has been excited i.e., the antisymmetric mode is infinitely damped, and that a warping operation has been applied to the symmetric mode. Following assessment of the mono-modal performance of the WFT, a bi-modal analysis was conducted. In the bi-modal analysis, the two modes were equally damped, and the raw signal was warped using the same titling convention as described previously. The phase velocity at the peak frequency component (200 kHz) for each of the warping modes was likewise used in conduct of the localisation.

**Signal-Averaging Validation**

The results of the second parametric study are presented below. Fig. 4 illustrates the reduction in SNR through application of the signal-averaging methodology for both raw and warped signals across all sensor spacings. As shown, the reduction in SNR showcases similar trends in both data sets with overall decreases for all sensor spacings.

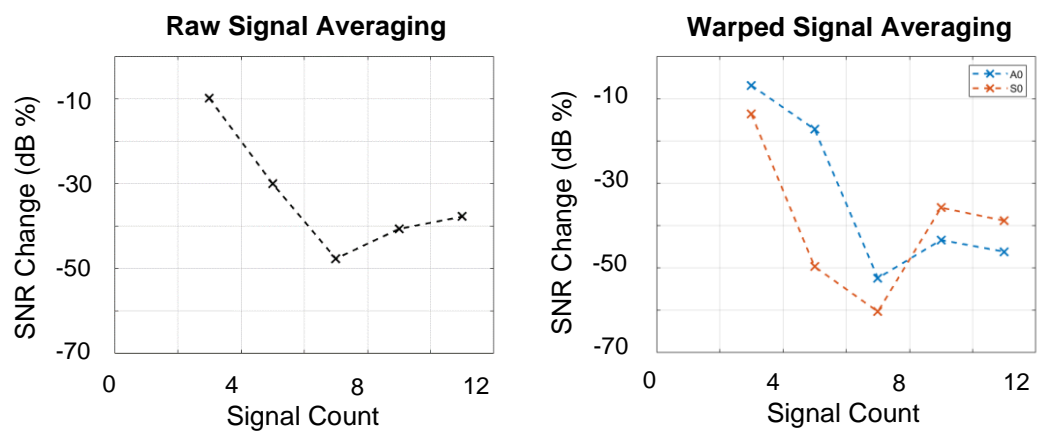

*Fig. 4: SNR reductions from the signal-averaging procedure.*

Initial results indicate that the signal-averaging procedure has been highly effective at reducing the SNR of both the raw and warped signals. However, further investigation has indicated that due to the bias of the cross-correlation, the underlying structure of the signals has been altered – a highly undesirable effect. Fig. 5 showcases an example of the signal-averaging methodology applied to both the raw and warped signals. Whilst cross-correlation has been largely successful at aligning the signals, some misalignments have resulted in changes to the signal structure. Finally, the results from the localisation are demonstrated in Table 2 below. As evident, extraction of the S0 mode using the signal-averaging methodology and WFT has greatly improved the capability to localise the emission when compared to a simple bandpass filtered approach. Whilst the absolute error remains relatively large, the DoA error is small, indicating that this approach has the capability to estimate the direction of the source with good accuracy.

*Table 2: Localisation results for the signal-averaged WFT S0 extracted mode and the raw bandpass signal.*

|  | Guess (mm) | Absolute Error (mm) | DoA Error (º) |
|---|---|---|---|
| **S0 Extracted Averaged Warp Signal** | (734.3, 739.2) | 372.3 | 0.1922 |
| **Raw Bandpass Signal** | (203.0, 97.97) | 1414 | 25.76 |

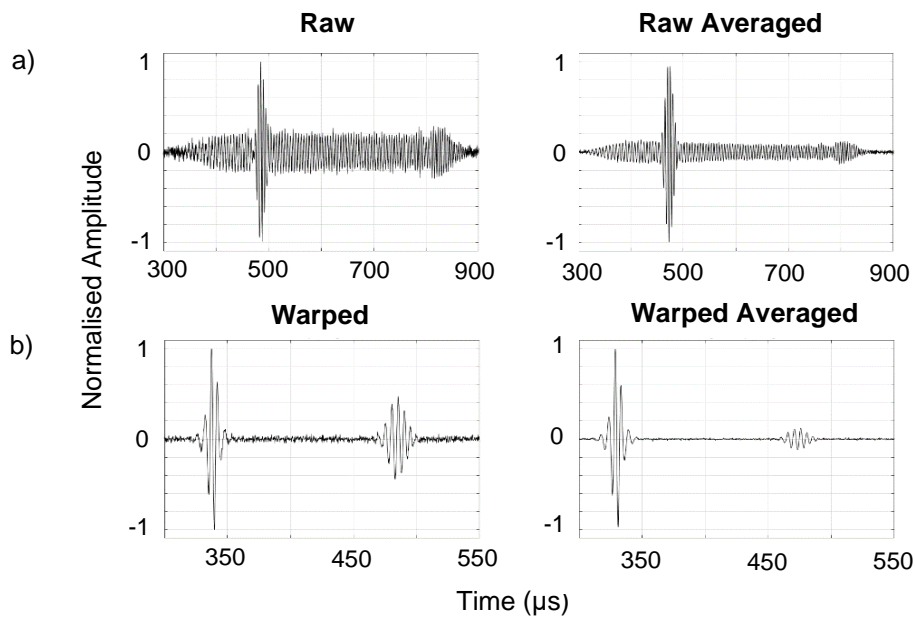

*Fig. 5: Results of the signal averaging upon the noised a) raw and b) warped with respect to the symmetric mode signals.*

## Conclusion

Ultimately, this study has indicated that the use of the WFT has good potential in the analysis of AEs for the purposes of Structural Health Monitoring (SHM). The WFT's ability to perform flexible tiling works well to separate multi-modal signals in highly dispersive regions. Moreover, such separation yields the possibility for mode extraction and independent localisation for improved performance. The WFT likewise presents the possibility to expand upon the PPA sensor array to allow for analysis of sparsely acquired signals using phased array algorithms. Further work is required to validate the WFT across a wider range of conditions, however, initial results are promising. The study of the recursive signal-averaging methodology likewise indicates that further testing is required but does produce some promising preliminary results. The combination of dispersion compensation with the signal-averaging method is a necessity as dispersion within raw signals is far too dominant to apply such techniques. Assessment of other time delay methods to improve upon signal alignment in artificial time-locking would prove beneficial.